\documentstyle[twocolumn,prl,aps,epsf,epsfig]{revtex}
\begin{document}
\twocolumn[\hsize\textwidth\columnwidth\hsize\csname
@twocolumnfalse\endcsname
\title{From HIV infection to AIDS:\\ 
A dynamically induced percolation transition?}
\author{Christel Kamp \cite{email} and Stefan Bornholdt}
\address{Institut f\"ur Theoretische Physik, Universit\"at Kiel,
Leibnizstra{\ss}e 15, D-24098 Kiel, Germany}
\maketitle
\date{today}
\begin{abstract}
The origin of the unusual incubation period distribution
in the development of AIDS is largely unresolved.
A key factor in understanding 
the observed distribution of latency periods,
as well as the occurrence of infected individuals not 
developing AIDS at all, is the dynamics of the
long lasting struggle between HIV and the immune system.
Using a computer simulation, we study
the diversification of viral genomes under 
mutation and the selective pressure of the immune system. 
In common infections vast spreading of viral genomes
usually does not takes place.
In the case of an HIV infection 
this may occur, as
the virus successively weakens the immune system
by depletion of $CD4^+$ cells.
In a sequence space framework,
this leads to a dynamically induced percolation transition, 
corresponding to the onset of AIDS.
As a result, we obtain 
the prolongated shape of the incubation period distribution, 
as well as a finite fraction of non-progressors that do not develop AIDS,
comparing well with results from recent clinical research.
\medskip \\ 
\end{abstract}
]
It is a well known empirical fact that incubation times of most diseases 
obey a lognormal distribution, only varying in their distributions' mean
 and dispersion factors. This has been verified for many single-exposure,
 common vehicle epidemics and is often referred to as ``Sartwell's model''
\cite{sartwell:1966,sartwell:1995}. Vice versa, this allows 
for estimates of exposure types as well as etiology of disease from an observed
incubation period distribution
\cite{glynn:palmer:1992,horner:samsa:1992,yi:glickman:1995}. 
Recently, the underlying dynamics that generate the incubation period
 distribution, as well as mechanisms that lead to deviations from the common 
distribution, have gained attention
\cite{philippe:1993,philippe:1994,philippe:2000}.
\newline
One of the most  prominent examples for a deviation from the lognormal
case is the distribution of waiting times between HIV infection 
(seroconversion) and the onset of AIDS, which is supported by data sets 
from various studies 
\cite{ukregister:1998,porter:darbyshire:1999,collaborativegroup:2000,cascadecollaboration:2000,thecascadecollaboration:2000,rki1998}. 
The divergence from lognormality, 
extraordinarily long incubation times,
and the occurrence of non-progressors (patients not developing AIDS)  suggest 
a more complex generating dynamics than observed in other infectious diseases. 
While much effort has been spent on parametric estimates of the 
incubation period distribution \cite{dangerfield:roberts:1999},
we here ask which are possible mechanisms of the underlying dynamics.
Any such attempt has to take into account the HIV-specific negative feedback
 to the host's immune system. While the immune system develops an
epitope-specific answer to HIV, as it does to any other antigenic invasion,
 it is weakened by HIV in a way 
that is not common to other viral infections. 
HIV targets the replicative machinery of $CD4^+$ cells which are 
depleted when viruses proliferate. 
$CD4^+$ cells as T helper cells are essential actors
within an immune response.
Therefore  HIV is able to globally weaken
 the host's resistance against antigens.
\newline
In earlier approaches,
the onset of AIDS has been associated
 with the passage of an antigenic diversity 
threshold in the framework of differential equation models
 \cite{nowak:may:1991,nowak:may:book}.
More recently, progress has been made to overcome the limitations of 
analytical models with respect to topological effects in the shape space of 
receptors and in physical space. Cellular automaton models
 have been defined and investigated that show the typical separation 
between the time scales of primary infection and the onset of AIDS
\cite{hershberg:solomon:2000,zorzenon:coutinho:2001}.
\newline
In this article we take an alternative approach and combine 
cellular automata with a sequence space framework in order to 
model typical characteristics of the time course of HIV infection.
In the following sections we will first define a framework to represent 
ordinary infections within the scope of percolation theory.
From there we will extend the model to describe the special case of 
HIV infection and discuss the  distribution of incubation periods.
Numerical simulations will be complemented by a stochastic model 
for the origin of the variety in incubation period distributions.
Finally we discuss our findings in the context of empirical data on
HIV survival.

\section{Percolation model of infection}\label{percolationsec}
Along the course of an infection one generally observes a diversification 
of viral genomes due to mutation and the selective
pressure of the immune system.
This co-evolutionary dynamics can be modeled within a
sequence space framework \cite{kamp:bornholdt:2002.1}.
Representing viral genomes by strings of length $n$, built up 
from an alphabet of length $\lambda$, we can describe their diversification
as spread in sequence space.
Analogously, let us assign a sequence to 
the respective immune receptor matching
the viral strain. Any string in sequence space is assumed
to represent a viral epitope, as well as its complementary immune
receptor. Thus each sequence is characterized by a viral and
an immunological state variable.
A mathematical framework to describe the dynamics in such a space can be found
in percolation theory \cite{stauffer:aharony:book} and in theories
for epidemic spreading, i.e. SIR models
\cite{hethcote:2000,moreno:vespignani:2001}. 
However, while those models apply cellular automata
to the interaction of organisms, we here apply the mathematical 
concept to modeling the populations of immune cells and viruses within 
one organism.
Adopting the notation of SIR models, we call a site in sequence space
susceptible, if it in principle can harbor a virus. 
It is denoted as infected, 
if the system contains a virus with an epitope motif
represented by the site's string. If a viral sequence meets immune 
response it is removed and the system is immunized against it.
In this case and in case a site in principle is inaccessible for a virus,
 it is called recovered (or removed). Aside from this,
two immunological states are distinguished.
An immune receptor shape may  
or may not be present within the immune repertoire.
We set up a system in which a site is inaccessible for viral 
sequences  with probability $D_0$ accounting for the fact that 
the viral genome is not arbitrary. 
In addition we introduce a probability of immunological 
presence at any site in sequence space $\rho_{is}(t)$ 
with $\rho_{is}(0)=\rho_0$.
This means that for sufficiently large systems
the initial density of recovered sites is 
$R(0)=D_0+\rho_0-D_0\rho_0$. Taking also into account the densities of 
susceptible sites $S(t)$ and of infected sites $\rho_v(t)$ one obtains 
the relation
\begin{equation}
S(t)+\rho_v(t)+R(t)=1 \qquad \forall t.
\end{equation}
As replication of viral and immunological entities is afflicted with 
copy fidelities $q_v<1$ and $q_{is}<1$, the system shows viral - and in
response immunological - spread in sequence space.
Introducing some viral strains into a so far unaffected system leads to
a dynamics that is modeled within the cellular automaton approach by
iterating the following steps:
\begin{enumerate}
\item Choose a random site.
\item If the site represents an active immune receptor
\begin{enumerate}
\item mutate any bit with probability $1-q_{is}$
\item if a new immunological strain is 
generated and the mutant matches an infected site
reset the site's viral status to recovered and assign the 
immunological state to be positive.
\end{enumerate}
\item If the site is infected
\begin{enumerate}
\item mutate any bit with probability $1-q_v$
\item if a new strain is generated and corresponds to a susceptible site 
the site gets infected.
\end{enumerate}
\end{enumerate}
A viral strain generates a specific mutant strain at Hamming distance $d$ 
(which is the number of differing bits)
with probability $\frac{(1-q_v)^d q_v^{n-d}}{(\lambda-1)^d}$ which 
will survive as long as it meets a susceptible site.
Equally an immunological mutant strain is originated with probability 
$\frac{(1-q_{is})^d q_{is}^{n-d}}{(\lambda-1)^d}$ under the condition that 
it coincides with an infected site. Otherwise we assume that the 
immunological mutant is not sufficiently amplified to establish a new strain.
\newline
Such a system shows two regimes of qualitatively different behavior. 
Below a percolation threshold depending on the above parameters the source of 
infection will stay negligible in size compared to the system size,
such that $R(\infty)=R(0)$ in the limit of infinite 
system size. 
Above the percolation threshold a virus will spread all 
over the system before it gets defeated. Accordingly $R(\infty)>R(0)$. 
\newline
To determine the threshold conditions within a mean field approach (``fully
mixed'' approximation), we introduce  the following system of differential 
equations
\begin{eqnarray}
\frac{dS}{dt}&=&-\sum_{d=1}^n{n\choose d}(\lambda-1)^d\frac{(1-q_v)^dq_v^{n-d}}{(\lambda-1)^d}\rho_vS\nonumber\\
&=&-(1-q_v^n)\rho_vS\\
\frac{d\rho_v}{dt}&=& -(1-q_{is}^n)\rho_{is}\rho_v+(1-q_v^n)S\rho_v\\
\frac{d\rho_{is}}{dt}&=&(1-q_{is}^n)\rho_v\rho_{is}\\
\frac{dR}{dt}&=&(1-q_{is}^n)\rho_{is}\rho_{v}
\end{eqnarray}
supplemented by the boundary conditions:
\begin{eqnarray*}
S(0)&\approx&1-R(0)\\
\rho_v(0)&\approx&0\\
\rho_{is}(0)&=&\rho_0\\
R(0)&=&D_0+\rho_0-D_0\rho_0.
\end{eqnarray*}
With $(1-q_{is}^n)\rho_{is}(t)>0$ for all $t$ and
$\rho_{is}(t)=R(t)-D_0+D_0\rho_0$ one can derive a relation between 
$S(t)$ and $R(t)$ 
\begin{eqnarray*}
\frac{dS}{dt}=-\frac{1-q_v^n}{1-q_{is}^n}\frac{\frac{dR}{dt}}{R-D_0+D_0\rho_0}S
\end{eqnarray*}
which yields
\begin{eqnarray}
S(t)\!\!&=&\!\!(1-D_0-\rho_0+D_0\rho_0)\!\left(\frac{\rho_0}{R(t)-D_0+D_0\rho_0}\right)^{\!\frac{1-q_v^n}{1-q_{is}^n}}\!\!\!\!\!
\end{eqnarray}
taking into account the above boundary conditions.
To evaluate conditions for the percolation threshold we can utilize 
\begin{eqnarray*}
R_{\infty}=R(\infty)=1-S(\infty)=1-S_{\infty}
\end{eqnarray*}
because in the stationary state any infected site will recover.
This leads to the following relation for $R_{\infty}$
\begin{eqnarray}\label{fpequation}
R_{\infty}\!\!&=&\!\!1-(1-D_0-\rho_0+D_0\rho_0)\!\left(\frac{\rho_0}{R_{\infty}\!-D_0+D_0\rho_0}\right)^{\!\frac{1-q_v^n}{1-q_{is}^n}}\nonumber\\
&=&\!\!f(R_{\infty}).
\end{eqnarray}
It is fulfilled for $R_{\infty}=D_0+\rho_0-D_0\rho_0=R(0)$ 
which means that no virus enters the system or at least cannot
gain macroscopic areas in sequence space.
However, the above equation has another solution if 
\begin{eqnarray*}
\frac{d}{dR_{\infty}}f(R_{\infty})_{|R_{\infty}=R(0)}>1,
\end{eqnarray*}
because $f(R(0))=R(0)$, $\lim_{R_{\infty}\to\infty}f(R_{\infty})=1$, $f(R_{\infty})<1\quad\forall R_{\infty}<\infty$. 
Evaluating the above condition leads to the result that an invading virus
can percolate in sequence space if
\begin{equation}\label{percolcondition}
\frac{1-q_v^n}{1-q_{is}^n}>\frac{\rho_0}{1-R(0)}.
\end{equation}
It is worthwhile to consider the case $q_{is}\rightarrow 1$ which implies that 
the above inequality holds for any $R(0)<1$. If the immune cells have 
vanishing mutation rates and accordingly lack adaptability, the virus 
percolates the sequence space in any case - unless immune  cells occupy any 
site in sequence space.
\newline
For $\frac{1-q_v^n}{1-q_{is}^n}=1$ we can explicitly determine the asymptotic 
density of recovered cells from the fixed point equation (\ref{fpequation})
that is solved by $R_{\infty}=1-\rho_0$ and 
$R_{\infty}=D_0+\rho_0-D_0\rho_0=R(0)$. 
With the additional constraint that $R_{\infty}\geq R(0)$ one can see 
that $R_{\infty}$ decays linearly with increasing $\rho_0$ until it is 
equal to $R(0)$ and the subcritical  regime is reached.
This is confirmed by computer simulations with various sets
of parameters. For the example of $D_0=0.5$, $q_v=q_{is}=0.95$, $n=15$,
$\lambda=2$  this leads to a critical immunological density $\rho_0^c=0.32$ 
(the theoretical value from equation (\ref{percolcondition}) 
is $\rho_0^c=\frac{1}{3}$).
\newline
Obviously in common infections the system is below the percolation threshold 
as an adequate immune response can defeat a viral attack before strains spread 
all over sequence space.
Nonetheless it is not unreasonable to assume 
that the immune system operates near the
percolation threshold as unnecessarily high immune receptor densities
$\rho_0$ involve competitive disadvantages.

\section{Percolation transition from HIV infection to 
the onset of AIDS}\label{hivmodel}
We are now in the position to extend our model to include HIV dynamics. 
An HIV model has to take care of characteristic
peculiarities of  HIV infections, 
i.e. the destruction of the immune system by the virus.
We consider this by extending the algorithm of section \ref{percolationsec}
by the following rule:
At any iteration step each viral strain is given
a chance to meet  a random immunological clone with
 probability $\rho_{is}(t)$ which thereafter is destroyed with probability $p$.
 If the affected site in principle is accessible for a viral strain the viral
status changes back to susceptible. 
We initialize the system near, but below the percolation threshold, 
which is the natural state of
a healthy immune system. As the system's qualitative behavior
shows to be insensitive to the specific choice of 
parameters we will in the following 
choose the parameter settings: 
$D_0=0.5$, $\rho_0=0.325$, $q_v=q_{is}=0.95$, $n=15$, $\lambda=2$.
\begin{figure}[h!]
\begin{center}
\begin{minipage}[h]{0.4cm}
\small{$\rho_v(t)$}
\end{minipage}
\begin{minipage}[h]{7.5cm}
\epsfig{file=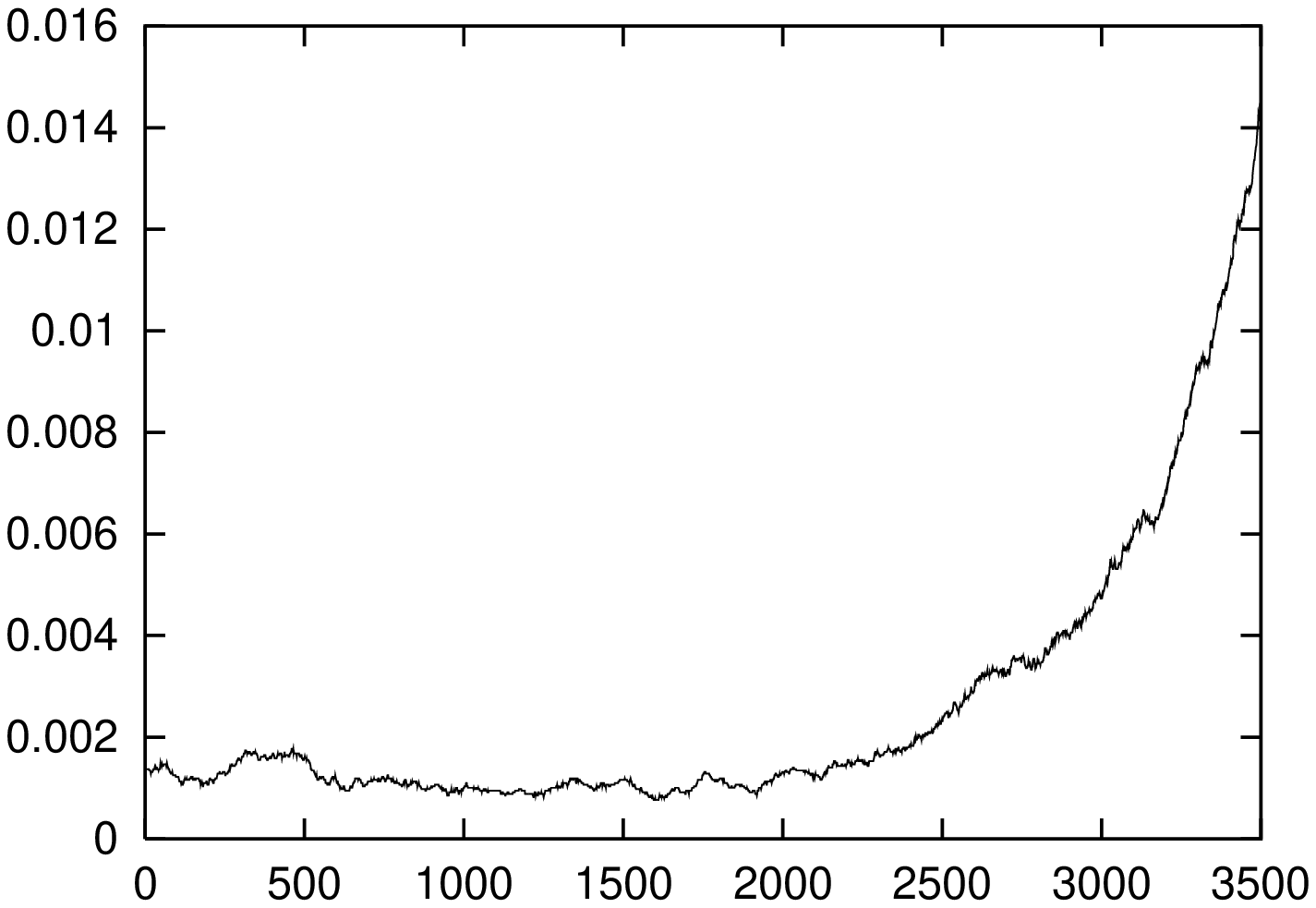,height=7.5cm,angle=270}
\end{minipage}
\newline
\small{$t$ [$1000$ Iterations]}
\caption{\label{vdiv}
Density of viral strains $\rho_v(t)$ in sequence space
under evolution of the system ($D_0=0.5$, $\rho_0=0.325$, $q_v=q_{is}=0.95$,
 $n=15$, $\lambda=2$, $p=0.0001$).
}
\end{center}
\end{figure}
\begin{figure}[h!]
\begin{center}
\begin{minipage}[h]{0.4cm}
\small{$\rho_{is}(t)$}
\end{minipage}
\begin{minipage}[h]{7.5cm}
\epsfig{file=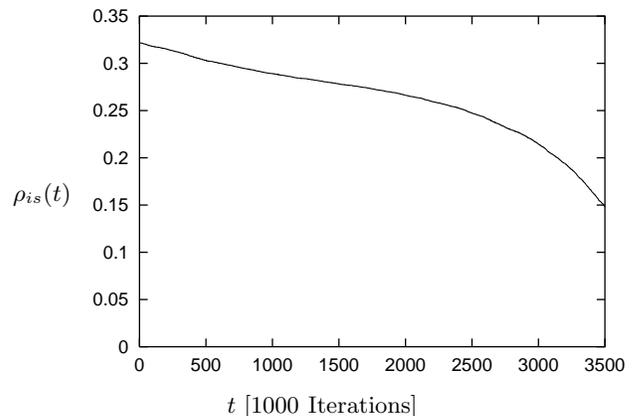,height=7.5cm,angle=270}
\end{minipage}
\newline
\small{$t$ [$1000$ Iterations]}
\caption{\label{cd4}
Density of immunologically active sites $\rho_{is}(t)$ 
in sequence space 
($D_0=0.5$, $\rho_0=0.325$, $q_v=q_{is}=0.95$, 
$n=15$, $\lambda=2$, $p=0.0001$).
Note the analogy to the decline in $CD4^+$ cells
under HIV infection.
}
\end{center}
\end{figure}

Figures \ref{vdiv} and \ref{cd4} show simulation results
for $p=0.0001$ exhibiting
characteristics typical to the course of disease
from HIV infection to the onset of AIDS.
One observes a drift of viral epitopes due to immune pressure 
as found in HIV-infected individuals
\cite{wei:shaw:1995,ganeshan:wolinsky:1997,yamaguchi:gojobori:1997,zhang:mayer:1997,phillips:mcmichael:1991,allen:watkins:2000,barouch:letvin:2002}.
Moreover, the simulations show fluctuations in the total number 
of actual strains, eventually sharply increasing which corresponds to
the onset of AIDS \cite{nowak:may:1991}. In parallel it is an empirical
 fact that the disease progresses with a depletion of $CD4^+$ cells
\cite{vergis:mellors:2000,karlsson:sodroski:1997,ganeshan:wolinsky:1997,endo:martin:2000}
which can be assumed to be accompanied by a loss in diversity of the 
immune repertoire as shown in figure \ref{cd4}. 
In this picture the immune system is successively weakened 
while fighting the viral attack 
and ultimately breaks down when the virus begins to percolate in sequence 
space. The virus dynamically drives the system from a 
subcritical regime above the percolation threshold. 
\newline
It will be interesting to investigate the distribution of waiting times
until percolation among systems only differing in their random 
initialization, which corresponds to the incubation period distribution.
\newline
To understand the generated distribution from a theoretical point of view
we have to take care of the stochastic nature of $\rho_v$ 
as seen in figure \ref{vdiv}. 
We assume that $\rho_v$ has a time dependent growth rate $r(t)$ that is 
superposed by noise and accordingly follows a generalized
geometric Brownian motion (cp. appendix).  
This process $\rho_v$ has a lower absorbing boundary for
 $\rho_v(t)=2^{-15}$ and converts into exponential growth after having passed
an upper point of no return $\rho_v^c$.
The first passage time distribution with respect to the upper boundary
corresponds to the incubation period distributions under investigation.
It is derived in the appendix and will be discussed in the context 
of simulation results and empirical HIV data in the following section.

\section{Results and Discussion}\label{resultsanddiscussion}
\begin{figure}[h!]
\begin{center}
\begin{minipage}[h]{0.4cm}
\small{Fract.}
\small{of}
\small{survivors}
\end{minipage}
\begin{minipage}[h]{7.5cm}
\epsfig{file=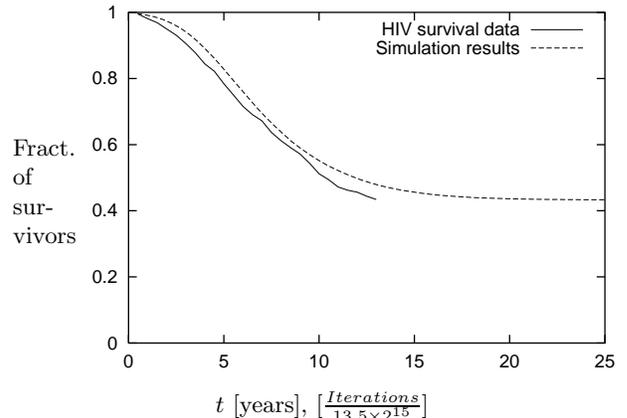,height=7.5cm,angle=270}
\end{minipage}
\newline
\small{$t$ [years], [$\frac{Iterations}{13.5\times2^{15}}$]}
\caption{\label{survival}
Comparison of the probability for HIV positives not yet to have developed AIDS
with a survival distribution generated by our simulations (after adequate 
renormalization of the time axis, $D_0=0.5$, $\rho_0=0.325$, $q_v=q_{is}=0.95$,  $n=15$, $\lambda=2$, $\rho_v(0)=0.0012$, $\rho_v^c=0.002$).
}
\end{center}
\end{figure}

We have run simulations as described in section \ref{hivmodel} for various
sets of parameters qualitatively leading to the 
same results for the time course
 of $\rho_v$ and $\rho_{is}$, as long as the system is initialized near but 
below the percolation threshold.
For the following discussion let us choose the parameter settings
  $D_0=0.5$, $\rho_0=0.325$, $q_v=q_{is}=0.95$, $n=15$, $\lambda=2$. 
The virgin system is infected within a ball
 that includes one and two bit mutants leading 
$\rho_v(0)=2^{-15}(1+{15\choose 1}+{15\choose 2})(1-D_0-\rho_0+D_0\rho_0)\approx0.0012$. A lower absorbing boundary of $\rho_v$ is given by $2^{-15}$ as 
less than one viral strain cannot exist.
Further evaluation of the simulations yields estimates
of $\rho_v^c=0.002$ where the virus begins to percolate.
Taking this together, we will be able to analyze the simulation results 
from the point of view of first passage time distributions 
(cp. appendix).
\newline
We have run simulations for various choices of $p$ mimicing viruses with 
different aggressiveness towards the immune system.
For $p$ as large as $0.005$ we hardly see any time period of struggle 
between the immune system and the virus leading to an immediate exponential
growth of $\rho_v$. The system shows very short incubation periods and 
vanishing probability of viral defeat. The distribution of incubation periods 
can then be approximated by a simple inverse Gaussian distribution.
Decreasing $p$ leads to longer incubation periods that correspond to 
periods of combat between virus and immune system as observed in figure 
\ref{vdiv}.
\newline
For further discussions we will focus on simulations with $p=0.0001$ as they
show a distribution of incubation periods that are in best accordance with
real data on HIV incubation periods. Nonetheless the theoretical framework  
as developed in the appendix will be equally 
applicable for arbitrary choice of $p$.
\newline
Figure \ref{survival} offers a comparison of a survival 
function generated by our cellular automaton model with the respective data
describing the probability that a HIV positive patient has not yet developed
 AIDS at time $t$ after seroconversion. The HIV data are taken from a
seroconverter study undertaken at the Robert Koch Institut
 within the CASCADE collaboration \cite{rkilink}.

\begin{figure}[h!]
\begin{center}
\begin{minipage}[h]{0.4cm}
\small{IPD}
\end{minipage}
\begin{minipage}[h]{7.5cm}
\epsfig{file=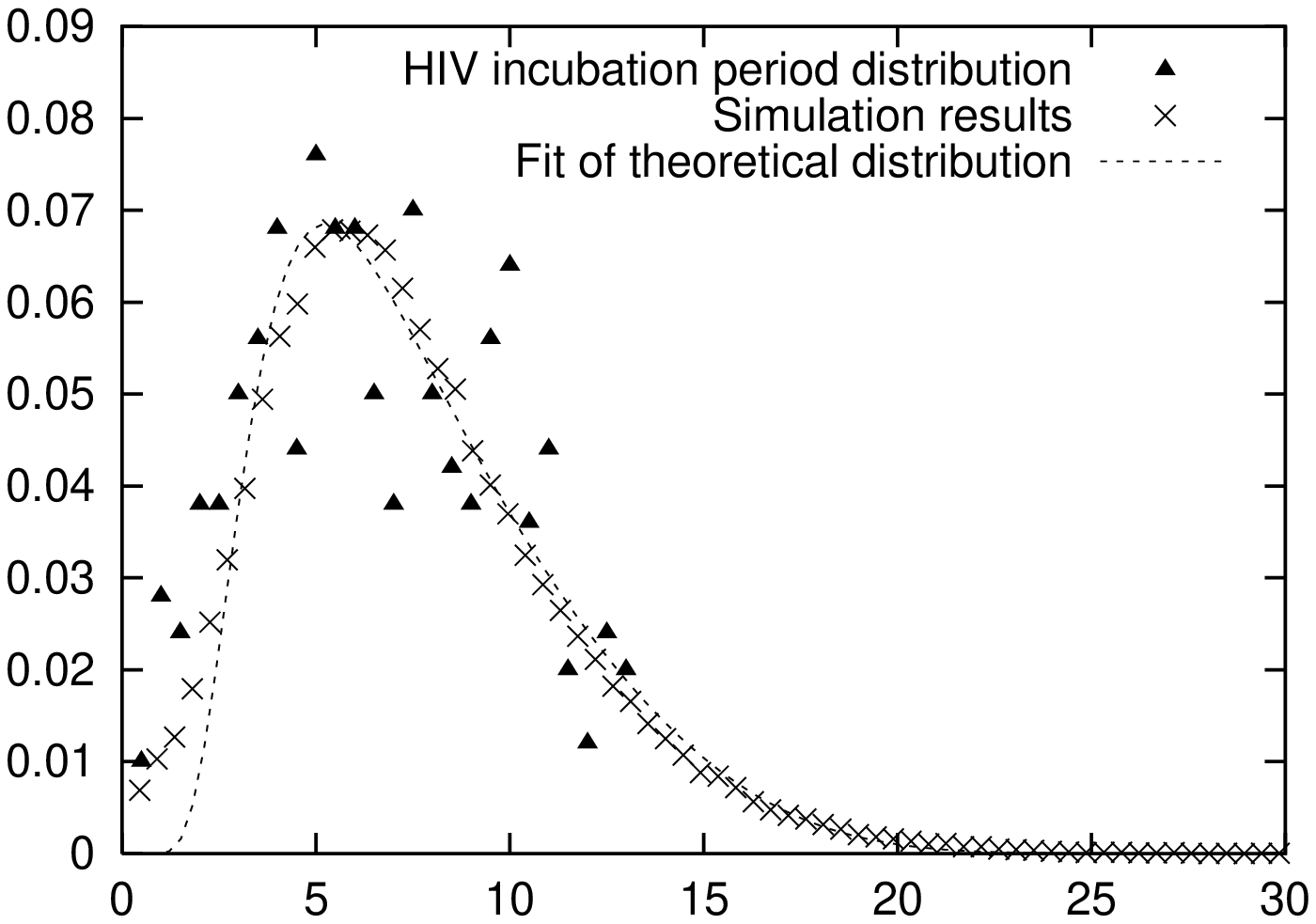,height=7.5cm,angle=270}
\end{minipage}
\newline
\small{$t$ [years], [$\frac{Iterations}{13.5\times2^{15}}$]}
\caption{\label{distrib}
Comparison of the incubation period distributions (IPD)
corresponding to figure \ref{survival} with the theoretical model 
with $r(t)=0.064-0.0092t$, $\sigma^2=0.0091$. 
}
\end{center}
\end{figure}

Figure \ref{survival} shows that the model 
reproduces main characteristics of the real system. The numerical simulations
result in a survival function that is similar to that observed
from HIV patients. In particular,
 they predict the occurrence of long-term survivors
as observed in reality and link it to a dynamical percolation mechanism.
We would like to emphasize that in this framework a quantitative
comparison of our model parameters with experimental data is not 
very meaningful.
However, any parameter setting that corresponds to a system that is initially 
below the percolation threshold and that is attacked with 
moderate aggressiveness (moderate values of $p$) will
show the same qualitative behavior. This demonstrates the robustness
of our model and ensures its applicability to
even larger sequence spaces than those simulated here.
\newline
Furthermore let us analyze the data in the light of the first passage time
distributions derived in the appendix. We have to specify
the functional from of the viral growth rate $r(t)$.
Different from the case of a very aggressive virus (large $p$), a constant
growth rate $r(t)=\mu>0$ does not fit the simulation results for 
viruses that are only moderately destructive (small $p$).
Therefore let us approximate $r(t)$ underlying the simulations 
by an expansion in powers of $t$ as
\begin{eqnarray*}
r(t)&=&\mu+\gamma t.
\end{eqnarray*}
Such a simple approach may not 
exactly reproduce the waiting time distribution but can show the origin of
its characteristics. This is exemplified by figure \ref{distrib} where the
incubation period distributions corresponding to the survival curves
shown in figure \ref{survival} are approximated by a first passage time 
distribution with $\mu=0.064$, $\gamma=-0.0092$ and $\sigma^2=0.0091$. 
This corresponds to the picture that the viral species initially is able
to establish new strains but that its opportunities for spreading in
sequence space are successively diminished. 
In many cases the virus nevertheless
is able to percolate sequence space if its suppression takes  
effect too slowly. 
This happens in a non-deterministic manner due to stochastic fluctuations 
corresponding to $\sigma^2>0$ and generates the observed incubation period 
distribution.

Limitations of the linear approximation become obvious with increasing
$t$. $r(t)$ is unbounded for negative numbers leading to
arbitrarily large
destruction of viral strains with time. As a consequence, the 
corresponding incubation period distribution shows an unrealistic
cut-off for large $t$. This disappears when considering
further terms in the expansion of $r(t)$ expanding the regime
of applicability of the theoretical model.
\newline
Describing the behavior of incubation periods within our model we 
can summarize that one observes an increase in waiting times 
before percolation and an enlarged fraction of cases where viral 
strains get totally extinct with decreasing $p$, i.e. less
aggressive viral strains.
This finds clear correspondence in real HIV statistics. $p$ is
a measure for the vulnerability of the immune
system under the attack of HIV. This virus manages its destructive
penetration into T helper cells ($CD4^+$ cells) not only by membrane
 fusion mediated by $CD4$ but generally needs an additional co-receptor
which is referred to as $CCR5$.
As almost all HIV strains rely on this mechanism for replication in T cells,
individuals who show a homozygous mutation leading to a non-expression
of the $CCR5$ receptor have proven to be resistant against HIV infection
\cite{lu:doms:1997}.
This is well in accordance with our model which for $p=0$ predicts
that no percolation will occur.
More recently it has been shown that also in individuals with
heterozygous genotypes a slower progression to 
AIDS can be observed. Moreover those patients have a $70\%$ reduced
risk to maintain the HIV infection and develop AIDS \cite{marmor:seage:2001}.
Therefore, already a reduction of $CCR5$ receptors 
on $CD4^+$ cells, making viral fusion more difficult,
improves the chance for prolonged or even total survival. 
This fits well with the predictions of the model for decrease in $p$.
\newline
Recent progress in 
vaccine research 
\cite{amara:robinson:2001,shiver:emini:2002,lifson:martin:2002}
further supports the model.
From the model's point of view, 
vaccination corresponds
to a local raise of immune receptors'
density $\rho_0$. This drives the system far below the percolation 
threshold and accordingly HIV will hardly manage to spread in sequence space.
\newline
In conclusion the above HIV/AIDS phenomenology
can be interpreted within
our cellular automaton model.
Prolonged survival as well as a finite fraction of non-progressors can be
traced back to the enhanced stability below the percolation transition
 in this framework.
Consequently, from the percolation model's point of view,
vaccination and receptor blocking
are encouraged
as efficient strategies to overcome an HIV infection.
\vspace{0.5cm}
\newline
C.K. would like to thank the Stiftung der Deutschen Wirtschaft 
for financial support.

\section*{Appendix:}\label{appendix}
\subsection*{First passage time distributions for geometric Brownian motion
between two absorbing boundaries}
Facing the stochastic nature of $\rho_v(t)$ we choose an ansatz
in the regime before the percolation transition
that expects a time dependent viral growth rate $r(t)$ of $\rho_v(t)$
which is superposed by noise.  
In terms of a stochastic differential equation this can be written as
\begin{eqnarray}
d\rho_v(t)
&=&r(t)\rho_v(t)dt+\rho_v(t)dB_t(0,\sigma^2)
\end{eqnarray}
with $B_t(0,\sigma^2)$ denoting Brownian motion with mean $0$ and variance
$\sigma^2t$. Within the Stratonovich interpretation 
\cite{oksendal:book} this equation leads to
\begin{eqnarray}
\rho_v(t)&=&\rho_v(0)e^{R(t)+B_t(0,\sigma^2)}\\
R(t)&=&\int_0^t r(t')dt'.
\end{eqnarray}
Accordingly $\rho_v$ is described by geometric Brownian motion
 that is locked between two absorbing
boundaries at $2^{-n}$ (less than one strain cannot exist) and an upper
critical concentration $\rho_v^c$ that leads to percolation of the virus.
This can be translated to Brownian motion $B_t(R(t),\sigma^2)$ (mean $R(t)$
and variance $\sigma^2t$) with $B_0=0$ and limited by
\begin{eqnarray*}
-a&=&\ln\left(\frac{2^{-n}}{\rho_v(0)}\right)<0\\
b&=&\ln\left(\frac{\rho_v^c}{\rho_v(0)}\right)>0.
\end{eqnarray*}
The probability density $p(x,t)$ describing the distribution
of the stochastic variable $B_t(R(t),\sigma^2)$ 
is determined by the following Fokker-Planck equation
\cite{honerkamp:book,gardiner:book}
\begin{eqnarray}
\frac{\partial p(x,t)}{\partial t}&=&
-r(t)\frac{\partial}{\partial x}p(x,t)
+\frac{\sigma^2}{2}\frac{\partial^2}{\partial x^2}p(x,t)\label{fpe1}\\
&=&-\frac{\partial}{\partial x}J(x,t)\\
J(x,t)&=&r(t)p(x,t)-\frac{\sigma^2}{2}\frac{\partial}{\partial x} p(x,t).
\label{J}
\end{eqnarray}
$J(-a,t)$ and $J(b,t)$ represent the contributions of the probability flow 
being absorbed at the boundaries $-a<0$ and $b>0$ at time $t$. 
In other words $J(b,t)dt$ is the probability that $B_t(R(t),\sigma^2)$ 
reaches $b$ for the first time in $[t,t+dt[$
 under the additional condition that it has not
yet met the absorbing boundary at $-a$. 
However, this means that $J(b,t)$ is equivalent to 
the first passage time distribution of the process
$\rho_v(t)$ with respect to the upper boundary $\rho_v^c$, again requiring
that it has not passed the lower absorbing boundary at $2^{-n}$.
Note that $J(b,t)$ represents a defective probability distribution 
in $t$ as the upper boundary is not reached with probability $1$.
\newline
Accordingly it remains to solve (\ref{fpe1}) with respect to the following 
initial and boundary conditions:
\begin{eqnarray*}
p(x,0)&=&\delta(x) \qquad \forall x\in[-a,b]\\
p(-a,t)&=&0 \qquad \forall t\\
p(b,t)&=&0 \qquad \forall t.
\end{eqnarray*}
Having the reflection principle in mind one can derive a solution under this
conditions as an adequate superposition of Gaussian distributions
\cite{honerkamp:book,gardiner:book,vankampen:book}. From this one can 
easily deduce using 
(\ref{J}) 
\begin{eqnarray}
J(b,t)\!\!\!\!\!&=&\!\!\!\!\!\frac{F(a,b,\sigma^2t)}{\sqrt{2\pi\sigma^2 t^3}}
e^{-\frac{(b-R(t))^2}{2\sigma^2t}}\\
F(a,b,\sigma^2t)\!\!\!\!\!&=&\!\!\!\!\!
\frac{e^{\frac{2b(a+b)}{\sigma^2t}}\!\!
\left(\!\!-a(1-e^{-\frac{2b(a+b)}{\sigma^2t}})
\!+\!b(e^{\frac{2a(a+b)}{\sigma^2t}}\!\!-1)\!\right)}
{e^{\frac{2(a+b)^2}{\sigma^2t}}-1}\nonumber\\
&\stackrel{a\rightarrow\infty}{\longrightarrow}&b.\nonumber
\end{eqnarray}
Obviously, in case of only one absorbing boundary (and $r(t)=\mu$,
$R(t)=\mu t$) we get the inverse Gaussian distribution 
as a well known solution 
for this special problem \cite{feller:book}.
A parameter setting of 
$D_0=0.5$, $\rho_0=0.325$, $q_v=q_{is}=0.95$, $n=15$, $\lambda=2$,
$\rho_v(0)=2^{-15}(1+{15\choose 1}+{15\choose 2})(1-D_0-\rho_0+D_0\rho_0)
\approx0.0012$ as discussed in section \ref{resultsanddiscussion} leads to
$a=3.7$ and $b=0.51$.

\end{document}